# Difficulties with Collapse Interpretations of Quantum Mechanics

Casey Blood
Professor Emeritus of Physics, Rutgers University
Sarasota, FL
Email: CaseyBlood@gmail.com

## Abstract

Quantum mechanics gives many versions of reality but we perceive only one. One potential explanation for this, the one considered here, is that the wave function collapses down to just one version. The experimental situation is briefly reviewed, with no evidence found for collapse. The theoretical position is also reviewed and found wanting. Collapse-by-observation schemes are logically untenable. A mathematical theory of collapse must be nonlinear, a significant departure from current quantum theory. In addition, the primary candidate theory, the GRW-Pearle model, requires instantaneous, non-local transmission of information. It also requires transmission of information across versions of reality, which is forbidden in current quantum mechanics. And there is no apparent physical quantity, such as particle number, that can provide the mechanism for collapse in all cases. Further, these aspects, which are in disagreement with current theory, seem to be necessary for *any* mathematical theory of collapse. The conclusion is that the outlook for mathematical collapse schemes is not encouraging.

PACS numbers: 03.65.-w, 03.65.Ta

## 1. Introduction.

When we look at the world around us, it certainly appears to be *the* physical world, upon which we all agree. But in quantum mechanics, the current, highly successful mathematical paradigm of physics, there is no unique physical world; instead there are many simultaneously existing versions of physical reality. This most peculiar state of affairs is nicely illustrated by the famous Schrödinger's cat thought experiment.

> A cat and a vial of cyanide are put in an enclosed box, while outside the box is a source of radiation and a detector. The detector is turned on for one minute. If it registers 100 or more counts, an electrical signal is sent to the box, the vial of cyanide is broken, and the cat dies. If it registers fewer than 100 counts, nothing happens and the cat lives. After the experiment is finished, the wave function of the system consisting of the detector, the cat, and an observer is the sum of two terms (actually, several terms if one includes all possible readings of the detector) which can be written schematically as

[det. reads more than 99 counts] [cat dead][obs. sees cat dead]
—*and*—
[det. reads less than 100 counts] [cat alive][obs. sees cat alive]

These two versions of reality—including two versions of the observer—*exist simultaneously* in quantum mechanics! How are we to reconcile this with our perception of a seemingly unique world? There are several possibilities. The first is to suppose that quantum mechanics is just plain wrong. But that simple 'out' will not do because quantum mechanics is so successful. It correctly predicts virtually all the properties of atoms, often to one part in a hundred million. It correctly gives the properties of the semiconductors used in all our many electronic devices. And along with a host of other successes, it correctly predicts many of the properties of elementary particles. In addition, it has never given a result that disagrees with experiment. So the first option, that quantum mechanics is wrong, is not really a possibility.

The second option is that the wave function of quantum mechanics is not the physical reality we perceive. Instead there is an actual, unique physical existence, made up perhaps of particles, which somehow conforms to just one of the versions of reality presented by quantum mechanics, and it is that physical reality which we perceive. But we have shown elsewhere [1] that that is also not a likely possibility. There is no evidence for it, and the very strict mathematics of quantum mechanics is not compatible with there being a reality underlying the wave function.

A third option, the one we will examine here, is collapse of the wave function/state vector. We can illustrate the basic idea by including the coefficients in the Schrödinger's cat wave function, so that it reads

$a_1$ [det. reads more than 99 counts] [cat dead][obs. sees cat dead]
—*and*—
$a_2$ [det. reads less than 100 counts] [cat alive ][obs. sees cat alive]

$$|a_1|^2 + |a_2|^2 = 1$$

The $a_1$ and $a_2$ denote the relative 'sizes' of the two possibilities, with their values following from the quantum mechanical equations. Now 'collapse' says that, for whatever reason, *either* $a_1 \rightarrow 1$, $a_2 \rightarrow 0$ so that the live cat version of reality has simply gone away—collapsed—and a dead cat is perceived; *or* $a_1 \rightarrow 0$, $a_2 \rightarrow 1$ and a live cat is perceived.  It is to be emphasized that *the conventional equations of quantum mechanics do not imply this collapse*; those equations imply that $a_1$ and $a_2$ stay the same size forever (see the appendix).  So if collapse occurs, its cause must be outside conventional quantum mechanics.  It is also to be emphasized that *there is no experimental evidence for collapse*.

There are two general categories of collapse.  The first, collapse by conscious perception, is considered in section 2, where it is shown that this type of collapse can essentially be ruled out.  The other category includes all models governed by mathematics.  As a preliminary to considering these, we outline a few of the properties of quantum states in section 3.  Then in section 4, we consider the most highly developed model of collapse, that of Ghirardi, Rimini, Weber [2], and Pearle [3], [4].  We find, however, that this model violates so many of the current restrictions on physical theories that, without further justification, it does not seem acceptable at this point.  Gravitational theories of collapse are briefly considered in section 5, but their current development is such that they are not currently viable.  The experimental situation is reviewed in section 6.  There is currently no evidence for collapse, but if collapse is assumed, the experimental results put restrictions on the theory.  Finally we note in the appendix that quantum mechanics must be modified so it is nonlinear in *any* mathematical model of collapse.  This prohibits decoherence from being the general cause of collapse.

## 2. Collapse by Conscious Perception.

In this conjectured method of collapse, the wave function stays uncollapsed until a human observer perceives the result.  (This apparently assumes each human being has a 'perceiving aspect' which is outside physical reality.)  So the cat is both dead and alive until 'consciously' observed by someone.  But there are, in my view, four major problems here.  The first is that, as we said, there is no evidence for this (or any other) type of collapse.  The second is that, because the equations of motion of conventional quantum mechanics do not alter the magnitudes of the coefficients $a_1$ and $a_2$, the act of observation must countermand those equations in a very specific, mathematical way (one coefficient goes to 1, the others go to 0).  This over-ruling of the mathematical equations by 'consciousness' simply doesn't seem acceptable to me.

The third problem has to do with the probability law.  How would collapse by conscious perception lead, after many runs of an experiment, to the $|a_i|^2$ probability law?

The final problem involves a variation of the Renninger experiment [5].  Do a Stern-Gerlach experiment on spin ½ silver atoms, with the source emitting atoms *randomly* at intervals of approximately 10 seconds.  The atoms in the – ½ state travel on one path and are detected by a nearby detector while the atoms in the + ½ state travel on another path and are detected by a more distant detector, with the difference in flight times between the two detectors being, say, .01 seconds. The observer sees only the reading of the $-\frac{1}{2}$ detector. The $+\frac{1}{2}$ detector has no visual readout, but it has an alarm that goes off 1 second after detecting a silver atom. When the alarm goes off, indicating detection on the $+\frac{1}{2}$ path, the $-\frac{1}{2}$ branch of the wave function must have been collapsed (because it wasn't observed) for 1.01 sec. But the observer *perceived nothing* before the alarm went off, so *the observer's perceptions could not have caused the wave function to collapse*. This simple reasoning, along with no rationale for the probability law, appears to definitively rule out collapse by conscious perception.

## 3. Properties of the Wave Function/State Vector.

Before we examine the Ghirardi-Rimini-Weber-Pearle (GRWP) mathematical collapse scheme, we need to review a few of the properties of the wave function/state vector. The six most important properties are discussed in detail in [1], so we will give an abbreviated version here.

**The wave function and the state vector.** The familiar Schrödinger equation of quantum mechanics is an equation for the wave function, so it is, on the least abstract level, the wave functions, rather than particles, which are the 'physical objects' in the mathematics of quantum mechanics. A useful conceptual picture of the wave function is that it is matter spread out in a mist or cloud of varying density. The Schrödinger equation determines the shape of the cloud, how it moves through space, and how it responds to other clouds corresponding to other 'particles.' The wave function of a macroscopic object like a cat or a human being, composed of billions of individual wave functions, is of course extremely complicated, but that does not prevent us from deducing its relevant general characteristics.

The actual state of affairs, however, is somewhat more abstract. Technically when discussing quantum mechanics, we should use the abstract 'state vector,' denoted by the ket notation, $|\ \rangle$, instead of the more concrete wave function. In the situations we shall discuss, however, the wave function description is essentially identical to the ket description, so it is permissible to use the less abstract wave function description.

### A. One and Only One Perception.

Schrödinger's cat is one example of a general property which follows from the agreement in all known cases between observation and the quantum mechanically predicted qualitative and quantitative characteristics of the versions of reality:

**A1.** Quantum mechanics gives many *potential* versions of reality, each represented by a ket.

**A2.** In any given instance, there is always—*always*—one and only one version that corresponds exactly to our physical perceptions. (This is an extremely far-reaching statement.)

**A3.** All observers agree on the perceived version.

The reason *why* properties **A2** and **A3** hold is a mystery. What we are investigating here is whether it is likely that they hold because there is collapse.

### B. Linearity. Different universes.

The most important property of quantum mechanics is that it is a linear theory. Because of this, when a state vector divides into a sum of different versions, each version evolves in time *entirely independently* of the other versions present. It is as if each version is in a different universe. (See the appendix for a proof and an expansion of this argument.)

**B1.** This has a bearing on why we perceive only one branch. The perception of different branches occurs in different universes that cannot communicate, and so we could never be communicably aware of the perception of more than one branch. The version of the brain perceiving one 'reality' could not communicate that perception to a version of the brain perceiving another 'reality.' Similarly, the division into different, non-communicating universes implies that two observers can never physically disagree on what they perceive.

**B2.** Note that linearity is not only used extensively in the calculations of virtually every result in quantum mechanics, but it is also necessary for establishing the particle-like properties of the state vector. Because of linearity and invariance properties, one can prove that mass, spin, and charge are properties of the state vector. If one takes linearity away, these theoretical results are in doubt.

### C. The Probability Law.

In the Schrödinger's cat experiment, the quantum mechanics of the nuclear decays will tell us that the 'sizes' (technically the norms, roughly the amount of cloud material) of the two possible wave function outcomes are different. The cat alive part of the wave function might have a size of 2/3, for example, while the cat dead part might have a size of 1/3 (so the wave function is written as $a_1$[cat alive]+$a_2$[cat dead], with $|a_1|^2$=2/3, $|a_2|^2$=1/3). If we do the Schrödinger's cat experiment many times (perhaps using a cat with many lives), there is an additional law in quantum mechanics which says that 2/3 of the time we will see a live cat, and 1/3 of the time we will see a dead cat.

More generally suppose the wave function contains *K* versions of reality, designated by *i* (*i*=1,…, *K*). Let the 'size' (norm) of version *i* be $|a_i|^2$, with the sum of the norms adding up to 1. Then the $|a_i|^2$ probability law says:

> If an experiment is run many times, a physical reality with characteristics corresponding to version *i* will be perceived a fraction $|a_i|^2$ of the time

This probability law is well-established, but the reason why it holds is not currently known.

> **C1. Consistency of the Probability Law.** Suppose we agree that the probability of event *i* is a function of the 'size' $|a_i|^2$ so that $p_i = f(|a_i|^2)$. Then there are any number of ways to show that the only functional form consistent with the mathematics of quantum mechanics is the conventional $f(|a_i|^2) = |a_i|^2$. The law $f(|a_i|^2) = (|a_i|^2)^2$, for example, would give inconsistent results. I will not pursue this point here, but I think **C1** strongly suggests that the origin of the $|a_i|^2$ probability law must, to a large extent, be *within* conventional quantum mechanics.

> **Note on collapse theories and particles.** In a collapse theory, there is the background assumption that physical existence consists of the state vector alone. This assumption itself depends on another assumption—that the state vector, by itself, without particles, is sufficient to account for all observations (property **A2**). So if one adheres to strict logic, one should show that quantum mechanics, by itself, can account for the photoelectric and Compton effects, and so on, before embarking upon the search for a mathematical collapse scheme. That is, strictly speaking, the reasoning in [1] is a prerequisite for collapse schemes.

## 4. The GRW-Pearle Mathematical Collapse Model.

### A. The Model.

In a mathematical theory of collapse, there are equations that govern the collapse. Because it is the most highly developed, we will primarily consider the mathematically elegant GRWP model of Pearle [3], [4], which builds on the work of Ghirardi, Rimini, and Weber [2]. A thumbnail sketch of the theory is that space is divided up into small volumes and a random potential energy is applied to all the particles (particle-like wave functions) in each of the volumes. If this random potential energy is chosen in accord with the probabilities assigned in the Pearle scheme, then one and only one state will be picked out (it's norm becomes very large compared to the others), with probability $|a_i|^2$.

More specifically, one starts from a Hamiltonian equation for the state vector $|\Psi,t\rangle$;

(1) $\quad \dfrac{\partial |\Psi,t\rangle}{\partial t} = -iH|\Psi,t\rangle + \left( \sum_n \left( \eta_n w_n(t) - \lambda \eta_n^2 \right) \right) |\Psi,t\rangle$

$H$ is the usual quantum mechanical Hamiltonian and is usually ignored in collapse theories (except when trying to find experimental tests of collapse). Space is divided up into fixed cubes of size on the order of $\alpha^3 = (10^{-5})^3$ cm$^3$. The $\eta_n$ are number operators; that is, when operating on a given version of reality, they give the number of particles in the $n^{\text{th}}$ cube. $\lambda$ is a frequency on the order of $10^{-16}$/sec, and $w_n(t)$ is the random part of the potential energy at time t acting on the particles in the $n^{\text{th}}$ volume element. $w_n(t)$ is related, in the Pearle formulation, to a random walk (or Brownian motion) parameter, $B_n(t)$ by

(2) $\quad w_n(t) = dB_n(t)/dt$

Let us now suppose the state vector is the sum of $K$ potential versions of reality,

(3) $\quad |\Psi,0\rangle = \sum_{k=1}^{K} a_k |k\rangle$

Then if we set $H=0$, Eq. (1) (for *given* $w_n(t)$) can be solved. Its solution is of the form

(4) $\quad |\Psi,t\rangle = \sum_{k=1}^{K} a_k \beta_k\big(B_1(t),\ldots,B_N(t),t\big)|k\rangle \equiv \sum_{k=1}^{K} \alpha_k(t)|k\rangle$

where the $\beta$'s are exponential functions of the $B$'s, and there are $N$ volume elements.

The probability for a given set of $w_n(t)$ in the Pearle model is

(5)
$$P(w_1(t),\ldots w_N(t)) = \left(\prod_{n=1}^{N} P_0(w_n(t))\right)\langle \Psi,t|\Psi,t\rangle$$
$$= \left(\prod_{n=1}^{N} P_0(w_n(t))\right)\left(\sum_{k=1}^{K} |\alpha_k(t)|^2\right)$$

where $P_0(w(t))$ is a white noise function. Roughly, Eq. (5) says that those random walks which increase the norm are favored (the norm does not stay constant in this model as it does in conventional quantum mechanics). Note that Eq. (5) does not hold at just one time; it holds at each instant (so the $B$'s are forced to 'walk' in time towards their more probable values).

Note also that this is a *nonlinear* model because Eq. (5) implies that the $w_n(t)$ depend (probabilistically) on the coefficients, $\alpha_k(t)$, so that $w_n(t)=w_n(t, \alpha_k(t))$ ). A different set of starting $a$'s will yield a different set of functions $w_n(t)$. Thus the Hamiltonian, through its dependence on the $w$'s, depends on the coefficients, so that Eqs. (1) and (5) together imply this is not a linear model.

> (One can see the nonlinearity more clearly if one imagines integrating the Hamiltonian equation on a computer, step by small time step. The choice of $w$'s or $B$'s at each step depends on the $\alpha$'s at that step, and the $\alpha$-dependent choice of $w$'s is put back into the Hamiltonian equation.)

It is Eq. (5), with its dependence on the values of the $a$'s in all the different versions, which makes this model radically different from current quantum mechanics.

## B. Difficulties with the Model.

There is nothing mathematically wrong with the model, but it has several questionable physics-related features which make it unlikely, in my opinion, that it is a 'correct' description of collapse.

> **(0) No evidence.** There is no experimental evidence for collapse by any mechanism (section 6).

> **(1) Form of the Hamiltonian.** We do not know what the physical source of the random potential energy is (Pearle's 'legitimization' problem [4]). The $w$'s must apparently correspond to some entirely new type of field because they make the equations nonlinear (section 4A and the appendix) and all current fields obey linear quantum mechanical equations of motion.
>
> > It is worth emphasizing how different the $w$'s are from what is conventionally encountered in quantum mechanics. Suppose we switch from the discrete $w_n(t)$ to a continuous $w(x,t)$ notation, and think of $w(x,t)$ as a fluctuating background field, the same in each version. That is quite acceptable, but it is not the whole story. $w(x,t)$ is determined in part by the $\langle \Psi | \Psi \rangle$ of Eq. (5), which represents an average over *all* versions! There is nothing like this—a quantity that is the same in all versions but whose value depends on an average of conditions in all the different versions—in current quantum mechanics.

> **(2) Use of particle number.** Pearle chose particle number (or the mass in a given cube) as the quantity which triggered collapse because differences in the positions of particles are the most obvious distinguishing feature of the different versions of reality. But it is unlikely that particle number (or the mass in a given cube) can lead to collapse in *all* situations (section 4C).

**(3) Nonlinearity.** Because Eq. (5) is nonlinear in the state vector, this theory of collapse is nonlinear. This is a huge departure from conventional quantum mechanics, where linearity is the most basic principle. More generally, *any* 'continuous spontaneous collapse' theory must be nonlinear (because of the results in the appendix), and giving up linearity is a significant change (section 3, point **B2**).

**(4) Coordination of random walks within a version.** The size of the volume elements is assumed to be about $10^{-15}$ cm$^3$. So if we have a typical detector volume of, say, 1 cubic millimeter, then there will be $10^{12}$ relevant volume elements, and hence $10^{12}$ different random walks.

Now if we think of the *B*'s as being produced by some physical process, then Eq. (5) says the processes for different volumes are not independent; the probabilistic choice of how $B_n$ changes from $t - \delta t$ to $t$ depends on the $B(t)$'s from all $10^{12}$ volume elements. What highly *non-local* physical mechanism could one imagine in which $10^{12}$ different random walks, each associated with a different volume element, are coordinated?

**(5) Coordination of random walks across different versions.** Further, the coordination is not just between elements from different volumes. It also goes *across versions* (the sum over values of *k* in Eq. (5)). In fact, no such cross-version mechanism is possible in current linear quantum mechanics because different versions are in different, non-communicating 'universes' (property **B** and the appendix). Coordination of random walks across versions can only happen in a theory that is nonlinear.

**(6) Coordination of random walks: instantaneous.** Because *t* occurs on both sides of Eq. (5), the process that determines each *B* at time *t* must have *instantaneous* knowledge of the values of all the other *B*'s at time *t*, both in the same version and in other versions. Physically, this seems most unlikely. (Actually, physically as opposed to mathematically, Eq. (5) looks inconsistent to me because the probability of each $B_n(t)$ cannot be known until all the $B_n(t)$'s are already chosen.)

**(7) Consistency with the $|a_i|^2$ law.** The very specific forms of Eqs. (1) and (5) were chosen so that the end result matched the $|a_i|^2$ probability law. There was no physical model or reasoning (apart from achieving the desired end result) which governed those choices. The net result is that one is deriving the $|a_i|^2$ probability law from another probability law, Eq. (5), whose origin is unknown and whose form is simply conjectured.

More generally, in *any* proposed mathematical collapse model, one must explain why the model should just happen to yield the only form—

the $|a_i|^2$ law—for the probability law that is consistent with quantum mechanics (property **C1**).

## C. Problem with particle number as hook.

The GRWP approach uses particle number as the 'hook' upon which to hang state vector reduction. If this is to work, then there must be a sufficiently large change in particle number in every situation. We will argue that this almost certainly is not true.

To start, we note that, as the collapse process proceeds, the ratio of the size of the collapsing part of the state to the non-collapsing part is approximately

(6) $\quad e^{-\lambda t (\Delta N)^2}$

so the approximate time for collapse is, say,

(7) $\quad \lambda t (\Delta N)^2 \cong 3$, or $t \cong 3/\left(\lambda (\Delta N)^2\right)$

As we said in section 4A, $\lambda \approx 10^{-16}$/sec. $\Delta N$ is the change in the number of particles in some particular volume, fixed in space, between the non-detecting and the detecting state. The size of the volume is taken to be $\alpha^3 = (10^{-5})^3$ cm$^3$. The simplest case to apply this idea to is that of a pointer which has one position for no detection and a different position for detection. In that case, $\Delta N$ is the number of particles in an appropriately located $10^{-15}$ cm$^3$ cube (the cube is filled with particles in one position and empty in the other), which is approximately $3 \times 10^{10}$ for normal densities. If we put these numbers into Eq. (7), we have a time of collapse of approximately $3 \times 10^{-5}$ sec, which is adequate (any collapse time less than $10^{-3}$ sec is physiologically undetectable).

But this situation—where the cube is filled with particles in, say, the non-detecting position and empty in the position corresponding to detection—is a best-case scenario. There are many situations where there is very little, if any, difference in the number of particles in any cube between the non-detecting and the detecting state:

> What if the readout device has a liquid crystal display? In any given $10^{-15}$ cm$^3$ cube, there is virtually no change in the number of particles between the state where it reads 0 and the state where it reads 1.

> What if the detection/readout device is a grain of film? Again, there is virtually no change in the number of particles (in a given cube) between the unexposed and the exposed states.

> And what if the detection/readout device is the human eye-brain system directly perceiving a flash of light? There is certainly an electrochemical

> change in various parts of the nervous system, but there is virtually no change in the *number of particles* (in any given cube) between the non-detecting and the detecting states.

In all these cases, because $\Delta N << 10^{10}$, the collapse time is unacceptably long. One possible way out is to adjust the values of $\alpha$ and $\lambda$, but the reduction in $\Delta N$ is so large in these and probably other cases that this almost certainly won't work. A second possibility is to use a different 'hook' (property) to trigger reduction. But at present, I can't imagine a single hook that would work in all situations. (Use of the mass in a cube is subject to the same criticism as the use of particle number.) So the 'hook' difficulty seems to me to be a serious problem for the GRWP collapse scheme.

## 5. Gravitational Models.

It has been proposed that gravity may be involved in collapse, either through a direct interaction as suggested by Penrose[6], [7] or through a model in which gravity leads to the fluctuating *w*'s [4], [8]. These proposals are consistent with the experimental results, which (if one assumes there is collapse) can be taken to imply that the coupling is proportional to the mass of the particle. I will not describe these models in detail but will instead emphasize that most of the difficulties of section 4B, as well as the restrictive arguments of the appendix, also apply to gravity-based collapse.

**The Penrose model.** There is a slightly different arrangement of matter in the different versions of reality of the state vector. And because the arrangement of matter affects the curvature tensor, and therefore the definition of spacetime, Penrose has suggested that the different versions create a 'tear' in the fabric of spacetime. This tear is postulated to produce a force between the various versions of reality, and if enough particles are involved, the force is sufficient to snap the different versions back into just one version.

The primary problem here is that the 'severe' orthogonality of equations (11) and (12) in the appendix will still hold, even though spacetime is different for the two versions. Thus if we stick with linear quantum mechanics, there can be *no interaction* between branches; that is, there can be no force which collapses the different versions into a single version. If the linearity of quantum mechanics is to be abandoned, the details of this radical change must be spelled out.

The second problem with the Penrose model is that there seem to be no random variables. Thus it is not clear how the process would lead to a random selection of states that obeys the $|a_i|^2$ law.

**The Stochastic Gravitational Model.** In this model, it is postulated, in essence, that there are fluctuations in mass distribution of the vacuum state.

These can be related to the fluctuating variables *w*. Thus the problem of item 1 in section 4B is somewhat circumvented (but at the cost of introducing additional assumptions on fluctuations of the vacuum), and one gains a feel for why the coupling should be proportional to the mass. But all the other problems remain. In particular, there is still an Eq. (5) or its equivalent, so the gravitational collapse model is nonlinear.

# 6. The Experimental Situation.

Our summary of the experimental situation will rely primarily on the review by Pearle [4].

**Interference experiments.** The most direct way to observe collapse would be to do an interference experiment, similar to those done on photons or neutrons. If the observed interference pattern did not follow that derived from standard quantum mechanics—if it disappears, for example—then this would imply that the wave function had collapsed to one branch or the other. This has been done with buckyballs having about 720 nucleons [9], and no deviation has been found from what is expected from standard quantum mechanics. This implies $\lambda < 10^{-6}$/sec so current interference experiments provide very little constraint on the theory (where the presumed value for $\lambda$ is on the order of $10^{-16}$/sec).

There have also been Superconducting QUantum Interference Device (SQUID) experiments in which approximately $10^9$ electrons took part [10]. The interference between two states of the $10^9$ electrons was just what quantum mechanics predicted, so there was no support for collapse. It is difficult to fit this into the GRWP scheme because there is no $\Delta N$. But still, when a state with $10^9$ particles shows no signs of collapse, that weakens one's confidence in the possibility that there is collapse.

**Decays and mass constraints.** The proposed GRWP form for the theory adds terms to the Hamiltonian (Eq. (1)) and so it can affect certain processes. In particular, it can lead to energy fluctuations that occasionally (*very* occasionally) cause the ejection of an electron from an otherwise stable atom. In one experiment [11], a chunk of germanium was monitored for a year to see if there were decays at a particular energy. The matrix element for the decay is substantial if electrons and nucleons couple equally to the collapse-producing fields, but it is essentially zero if the coupling is proportional to the mass. The negative results (no decays) are *consistent with no collapse*. If collapse is assumed, the results lead to the constraint

(8) $\quad 0 \leq \dfrac{\alpha_{elec}}{\alpha_{nuc}} \leq \dfrac{13 m_{elec}}{m_{nuc}}$

where the $\alpha$'s are the coupling constants (and $m_{elec}/m_{nuc} \cong 1/2,000$). That is, the results are consistent with the coupling being proportional to the mass, but completely inconsistent with an equal coupling for electron and nucleon.

In another experiment that yields information on the possibility of collapse, the smallness of the disagreement between theory and observation in the Sudbury experiment on solar neutrinos [12] is *consistent with no collapse*. But if we assume collapse, it gives a constraint on coupling constants for neutrons and protons;

(9) $\quad \dfrac{\alpha_n}{\alpha_p} = \dfrac{m_n}{m_p} \pm 4 x 10^{-3}$

This, too, is consistent with a coupling proportional to the mass.

To date, the mass constraints are the only serious constraints on GRWP theories, but they are significant. It apparently implies either that the coupling is proportional to mass (so there would be no coupling to photons and very little to electrons) or that the coupling is only to nucleons (in which case the random field would have internal symmetry group properties).

## 7. Summary and Conclusion.

The problem is to explain why, when quantum mechanics gives many versions of reality, we perceive just a single version and we all agree on the version. The potential explanation investigated here is that the state vector collapses down to just one version.

On the experimental side, there is no evidence whatsoever for collapse.

There are two broadly different 'theoretical' mechanisms proposed for collapse. The first is to assume it is caused by conscious perception. Logic—how to account for the probability law, and the Renninger experiment—is strongly against this possibility, virtually ruling it out.

The second type of mechanism is to have a mathematically governed collapse process. The GRWP model admirably solves the *mathematical* problem of collapse—how to get ($a_1, a_2, \ldots, a_i, \ldots, a_K$) to evolve to (0, 0, …, 1, …,0) a fraction $|a_i|^2$ of the time. But it encounters substantial problems when one looks at the physics implied by its mathematics.

- First, it appears that the use of particle number as the hook upon which to hang the state vector reduction process cannot be made to cause collapse in all cases.
- Second, the theory must be nonlinear. This is a major departure from current quantum mechanics, where linearity is the primary property.
- Third, the coordination between random processes located in volumes separated by macroscopic distances is instantaneous and non-local.

•Fourth, the coordination of random processes goes across versions of reality, which is strictly forbidden in current quantum mechanics.
•Finally, there is no justification for the very specialized forms of the Hamiltonian, Eq. (1), and probability law, Eq. (5), used to produce collapse, except that they give the correct answer.

In my opinion, the GRWP model, although it is ingenious and is the only well-developed mathematical model of collapse we have, does not lend strong support to the conjecture that collapse occurs. More generally, the difficulties with the GRWP model extend to the whole class of 'continuous spontaneous collapse' models, so the prospects for constructing an acceptable theory of collapse are, in my opinion, not encouraging.

# Further Work.

This is the second in a series of three articles. The first [1] showed that there is no evidence for particles and it is quite unlikely they exist. In this article, we noted that there is no evidence for collapse, and that the construction of an acceptable theory of collapse encounters substantial difficulties. It will be shown in the third article that if there are no particles, and if there is no collapse, then awareness (of one version of the wave function) cannot be based in the physical brain.

# Appendix.
# There Can Be No Collapse in a Linear Theory if the $|a_i|^2$ Probability Law Holds.

We will show that there cannot be a *general* cause for collapse in a linear theory of quantum mechanics. This result is demonstrated using a particular example, but the same arguments would hold whenever any 'single-particle' wave function (except perhaps that for a photon) simultaneously takes at least two different paths. Since we are dealing with low-energy physics here, we are justified in using wave function language rather than state vector language.

**1. Different branches are in different universes in a linear theory.** Fire a spin ½ silver atom into a Stern-Gerlach device. There will be one 'trajectory' traced out by the +½ states and another traced out by the –½ states. Let $\Omega_+(t)$ be the time-dependent three dimensional volume where the silver atom wave function is non-zero on the +½ trajectory and $\Omega_-(t)$ the corresponding volume on the –½ trajectory. These regions are well-defined even after the two branches of the silver atom wave function hit the detectors, and they are non-overlapping after the wave function of the silver atom clears the magnet.

Now consider the Hamiltonian equation for the full wave function (including both versions of reality) describing the silver atom and the detectors;

$$(10) \quad i\frac{\partial(a_+\Psi_+(x_s,\{x_d\})+a_-\Psi_-(x_s,\{x_d\}))}{\partial t} = H(a_+\Psi_+(x_s,\{x_d\})+a_-\Psi_-(x_s,\{x_d\}))$$

where the $a$'s give the relative sizes of the two versions, $x_s$ is the coordinate of the silver 'atom' and $\{x_d\}$ represents the coordinates of all the 'atoms' in the detectors (plus the observer if one wishes). $\Psi_+$ denotes the branch of the wave function when $x_s$ is in region $\Omega_+$, and $\Psi_-$ the branch when $x_s$ is in region $\Omega_-$. Because the two regions do not overlap, we have

$$(11) \quad \Psi_+(x_s,\{x_d\}) = i\partial_t \Psi_+(x_s,\{x_d\}) = H\,\Psi_+(x_s,\{x_d\}) = 0 \text{ when } x_s \text{ is in region } \Omega_-$$

$$(12) \quad \Psi_-(x_s,\{x_d\}) = i\partial_t \Psi_-(x_s,\{x_d\}) = H\,\Psi_-(x_s,\{x_d\}) = 0 \text{ when } x_s \text{ is in region } \Omega_+$$

> (Note: The $H$ in $H\Psi_+(x_s,\{x_d\})$ does not change the location of the silver atom. Thus $H\Psi_+(x_s,\{x_d\})$ in Eq. (11) is a function of $x_s$ which is 0 when $x_s$ is not in $\Omega_+$.)

Eqs. (10), (11) and (12) imply the wave functions for the two branches obey their own separate equations of motion,

$$(13) \quad i\partial_t \Psi_+(x_s,\{x_d\}) = H\,\Psi_+(x_s,\{x_d\}) \quad \text{(relevant when } x_s \text{ is in region } \Omega_+)$$

$$(14) \quad i\partial_t \Psi_-(x_s,\{x_d\}) = H\,\Psi_-(x_s,\{x_d\}) \quad \text{(relevant when } x_s \text{ is in region } \Omega_-)$$

(and the Hamiltonian equation is irrelevant (0=0) when $x_s$ is in neither region.) Because they obey their own separate equations, and because $H$ is independent of the wave function in a linear theory, the evolution of $\Psi_+$ is entirely independent of what happens on the $\Psi_-$ branch (and vice versa).

**2. Non-unitary Hamiltonian.** If $H$ is unitary, as it is in conventional theory, Eqs. (13) and (14) tell us the norms of the two branches stay the same forever. So there can be no collapse in this Stern-Gerlach example if one sticks to a linear, unitary theory. But what would happen if we kept the Hamiltonian linear but allowed it to be non-unitary? Suppose in particular that some part of the Hamiltonian can change the norm of the wave function, with the change depending on the random variables $w$, where the choice of $w$'s is independent of the coefficients. Then we have for the general linear case

$$\Psi(t,w) = U(t,w) \sum_{k=1}^{K} a_k \Psi_k(t=0)$$

(15)
$$= \sum_{k=1}^{K} a_k U(t,w) \Psi_k(t=0)$$

$$= \sum_{k=1}^{K} a_k \beta_k(t,w) \Psi_k(t=0)$$

where the conjectured Hamiltonian is presumed to have no effect on the 'shape' of the wave function (and we have ignored the non-collapse part of the Hamiltonian). In this case, it might happen that after some time has elapsed,

(16) $\quad \dfrac{\beta_i^2(t,w)}{\sum \beta_k^2(t,w)} \to 1, \ t \to \infty$

so we end up with a system which has collapsed to state $i$. Thus linearity does not rule out collapse.

    **3. The $|a_i|^2$ probability law requires collapse theory to be nonlinear.**
But now suppose the collapse follows the $|a_i|^2$ probability law. Then for large $t$ we must have Eq. (16) holding a fraction $|a_i|^2$ of the time and the same equation with 1 replaced by 0 holding the rest of the time. Thus to satisfy the $|a_i|^2$ probability law, the average of the ratio must obey

(17) $\quad \langle \dfrac{\beta_i^2(t,w)}{\sum_{k=1}^{K} \beta_k^2(t,w)} \rangle = \int dw P(w) \dfrac{\beta_i^2(t,w)}{\sum_{k=1}^{K} \beta_k^2(t,w)} = |a_i|^2, \ t \to \infty$

where $P$ is the probability of a certain set of $w$'s and the integral over all possible $w$'s is suitably defined. If $P(w)$ is independent of the coefficients, this cannot hold. (To see this, take the derivative of the second and third terms in Eq. (17) with respect to $|a_i|^2$. The derivative of the second term gives 0—no dependence on the coefficients—while the derivative of the third gives 1.) Thus to satisfy Eq. (17), $P(w)$ must depend on the coefficients (as in Eq. (5), for example).

    But this implies the reasoning of section 4A holds: At each step in the integration of the equations of motion, the choice of $w$'s dictated by $P(w) = P(w; \alpha_1, ..., \alpha_K)$ depends on the coefficients. And then the dependence of the Hamiltonian on the $w$'s gives an indirect dependence of the Hamiltonian on the coefficients, so that different $a_i$'s will produce different $w$'s. This makes the equation of motion nonlinear, in contrast to current quantum mechanics.

**4. Long-range Hamiltonian.** Eqs. (11-14) prohibit interactions between versions even if the Hamiltonian is very long-ranged. To see this more explicitly, note that $H$ contains no information about the location of the silver atom (or any other atom). Thus there can be no information in $H\Psi_+$ concerning the conditions—locations of atoms and so on—when $x_s$ is in $\Omega_-$. This implies the potential energy in $H\Psi_+$ depends on the arrangements of the 'particles' in the detectors when $x_s$ is in $\Omega_+$ (with a similar statement for $H\Psi_-$). So even if there is a long-range interaction, there can be no energy of interaction,

(18) $\quad \langle \Psi_-(t) | H | \Psi_+(t') \rangle = 0$,

and hence there can be no force or influence of any kind of one version on another in a linear theory. Because of this, once the split occurs, no version of reality can change its energy. (More generally, the isolation of versions implies that all conservation laws apply to each version *separately*.)

**5. Decoherence. The environment cannot be the general cause of collapse.** It has been suggested that the interaction with the environment causes—in all cases—the collapse to just one version. There may indeed be cases where the environment causes collapse, but the above argument shows this cannot be true in all cases. Nothing within linear quantum mechanics—including the environment—can lead to a collapse process that obeys the $|a_i|^2$ probability law in the Stern-Gerlach experiment.

> (One might try, in conventional quantum mechanics, having the wave functions of 'the environment'—including, perhaps, the vacuum state—somehow generate the random variables $w$. But section 3 of this appendix shows that will not work because, in conventional quantum mechanics, there can be no coordination between branches.)

Note that each version constitutes a different version of the *whole physical universe*, including 'the environment,' even though a good deal of the 'background' may have the same wave function in different versions.